\documentstyle[aps,twocolumn,epsf]{revtex}

\begin{document} 

\title{Universality in Three Dimensional Random-Field Ground States}
\author{A.~K.~Hartmann$^*$ and U.~Nowak$^{\dagger}$}
\address{$*$ Institut f\"{u}r theoretische Physik, Philosophenweg 19,
  69120 Heidelberg, Germany\\
  e-mail: hartmann@philosoph.tphys.uni-heidelberg.de\\   
  $^{\dagger}$ Theoretische Tieftemperaturphysik,
    Gerhard-Mercator-Universit\"{a}t-Duisburg, 47048 Duisburg, Germany\\
  e-mail: uli@thp.uni-duisburg.de 
}

\date{July 8, 1998}
\maketitle

\begin{abstract}
  We investigate the critical behavior of three-dimensional
  random-field Ising systems with both Gau\ss{} and bimodal
  distribution of random fields and additional the three-dimensional
  diluted Ising antiferromagnet in an external field. These models are
  expected to be in the same universality class. We use exact
  ground-state calculations with an integer optimization algorithm and
  by a finite-size scaling analysis we calculate the critical
  exponents $\nu$, $\beta$, and $\bar{\gamma}$. While the random-field
  model with Gau\ss{} distribution of random fields and the diluted
  antiferromagnet appear to be in same universality class, the
  critical exponents of the random-field model with bimodal
  distribution of random fields seem to be significantly different.
\end{abstract}



Above two dimensions, the ferromagnetic random-field Ising model
(RFIM) is long-range ordered for low temperatures and small random
fields as was proven by Imbrie \cite{imbrie} and also by Bricmont and
Kupiainen \cite{bricmont}. For larger fields the system develops a
frozen domain state \cite{imry} which has been shown to have a
complex, fractal structure \cite{esser}. It is now widely believed
that there is a second order phase transition from the ordered to the
disordered phase in appropriate dimensions although in three
dimensions a complete set of values of the critical exponents
fulfilling the predicted set of scaling relations
\cite{bray,nattermann,villain} could still not be established.
E.~g.~the value of $\alpha$ is still controversially discussed
\cite{nowak}.

For the replica-symmetric mean-field solution \cite{aharony} it was
found that the critical behavior of the RFIM depends on the kind of
distribution of random fields. Later, also for random-field systems on
the Bethe-lattice \cite{swift1} it was demonstrated that the critical
behavior depends on the distribution of random fields. Two recent
letters \cite{swift2,sourlas} were addressed to the question whether
this is also true in lower dimensions. Swift et al.  \cite{swift2}
found clearly different critical behavior for random-field systems
with a Gau\ss-distribution (G-RFIM) on the one hand and a bimodal
distribution (B-RFIM) on the other hand in four dimensions. They could
not find a clear distinction in three dimensions. Here, it were
Angl\`{e}s d'Auriac and Sourlas \cite{sourlas} who found differences
for the critical behavior of the two systems mentioned above.
Especially the values of the correlation length exponent $\nu$ they
found to be significantly different.

The most prominent experimental realization of a RFIM is often
asserted to be the diluted Ising antiferromagnet in a field (DAFF)
(for an overview see \cite{kleemann,belanger}).  This system is
thought to be in the same universality class as the RFIM
\cite{fishman,cardy} but if the concept of universality is violated
for random-field models the question arises what values the critical
exponents of a DAFF have.  Therefore, in this work we investigate
these three types of random-field systems mentioned above in three
dimensions and we determine at a time three of the critical exponents,
$\nu$, $\beta$ and $\bar{\gamma}$ numerically in order to test if
there is possibly a violation of universality. Especially the DAFF is
examined the first time in this way at all.

The Hamiltonian of the RFIM in units of the exchange coupling constant
is
\begin{equation}
  H = - \sum_{\langle i, j \rangle} \sigma_i \sigma_j - \sum_i B_i \sigma_i.
\end{equation}
The first sum is over the ferromagnetic nearest-neighbor interactions
and the spin variables $\sigma_i$ are $\pm 1$. The random
fields $B_i$ are taken either from a Gau\ss -probability distribution
$P(B_i) \sim \exp(-(B_i/\Delta)^2/2)$ or from a bimodal distribution
$B_i = \pm \Delta$.  In either distributions $\Delta$ scales the
strength of the random field.

The corresponding Hamiltonian of the DAFF is
\begin{equation}
  H = \sum\limits_{\langle i, j \rangle} \epsilon_i
  \sigma_i \epsilon_j \sigma_j - \Delta \sum\limits_i \epsilon_i \sigma_i
\end{equation}
where we have now an antiferromagnetic nearest neighbor coupling and
the $\epsilon_i = 0,1$ represent the dilution. The homogenous magnetic
field $\Delta$ breaks the antiferromagnetic long-range order and in
connection with the dilution it acts as random field
\cite{fishman,cardy}.  Note, that for the DAFF the value of the
critical $\Delta$ depends on the dilution of the system (see also
\cite{nowak2} for a sketch of the phase diagram of the DAFF).

As was shown by renormalization group arguments \cite{bray,berker},
the three-dimensional RFIM has a zero temperature fixed point at a
finite value $\Delta_c$ of the random-field width and the
temperature $T$ is an irrelevant variable. Hence, we can use exact
ground-state calculations to investigate the critical behavior of our
systems at zero temperature. 

For our numerical investigation we used a simple cubic lattice with
periodical boundary conditions and linear lattice sizes varying from
$L = 10$ to $L = 80$ for the RFIMs and from $L = 20$ to $L = 120$ for
the DAFF.  We used well known algorithms from graph theory
\cite{swamy,claibo,knoedel} to calculate the ground state of a system
at given field $\Delta$.  The calculation works by transforming the
system into a network \cite{picard1}, and calculating the maximum flow
in polynomial time \cite{traeff,tarjan}.  \footnote{Implementation
  details: We used Tarjan's wave algorithm together with the heuristic
  speed-ups of Tr\"aff. In the construction of the {\em level graph}
  we allowed not only edges $(v, w)$ with level($w$) = level($v$)+1,
  but also all edges $(v,t)$ where $t$ is the sink. For this measure,
  we observed an additional speed-up of roughly factor 2 for the
  systems we calculated.}

This method works only for systems without bond-frustration, so that
spin glasses cannot be treated in this way. For most of those systems
only algorithms with exponential time complexity are known, for
example the Branch-and-Cut method \cite{simone95,simone96}. For
readers interested in the field we give some additional informations:
Only for the special case of the two-dimensional spin glass with
periodic boundary conditions in no more than one direction and without
external field also a polynomial time algorithm is known
\cite{barahona82b}.  For the general case the simplest method works by
enumerating all possible states and has obviously an exponential time
complexity. Even a system size of $4^3$ is too large.  Branch-and-Cut
works by rewriting the problem as a linear optimization problem with
an additional set of inequalities which must hold for the solution.
Since not all inequalities are known a priori the method iteratively
solves the linear problem, looks for inequalities which are violated,
and adds them to the set until the solution is found. Since the number
of inequalities grows exponentially with the system size the same
holds for the computation time of the algorithm. Here only small systems
up to $8^3$ are feasible . For the spin-glass problem approximation
methods like combinations of Cluster-exact approximation \cite{alex2}
and genetic algorithms \cite{michal92,pal96} are more efficient: true
ground states \cite{alex_stiff} up to size $14^3$ can be calculated
\cite{alex_sg2}. The basic idea of Cluster-exact approximation is to
build sub-clusters of spins which exhibit no bond-frustration. For
these sub-clusters the graph-theoretical methods used here can be
applied, which leads to a decrease of the energy in the total system.
Genetic algorithms work by minimizing many configurations of a system
in parallel, keeping only those which have lower energies and creating
new configurations by combining already existing configurations and
flipping some spins randomly.

We now turn back to the RFIM and the DAFF.  All degenerate ground
states of the system are given \cite{picard2} by a set of clusters and
a binary relation defined on it. Each cluster is a set of
antiferromagnetically (DAFF) respectively ferromagnetically (RFIM)
ordered spins.  These spins are not necessarily spatially connected.
Two of the clusters hold the spins which have in all degenerate ground
states always the same orientation. The relation describes the
conditions which must hold between the orientations of the other
clusters in different ground states. Using this description all
degenerate ground states can be analyzed.  Since all systems have a
finite number of spins the ground state is a stepwise constant
function of the field and, hence, this holds for the measurable
quantities as well. The steps occur whenever a cluster of spins flips
it orientation. In \cite{alex_daff2} it was shown that for the DAFF
more than 95\% of the spins do not contribute to the degeneracy. For
the B-RFIM even 98\% of the spins are frozen in different ground
states.  This is true for all fields $\Delta$ except of the finite
number of fields where the ground state changes, i.~e. a jump in a
measured quantity of a single system occurs.

From the spin configurations of the ground state, we can calculate the
magnetization $m = \frac{1}{L^3}\sum_i \sigma_i$ (respectively
staggered magnetization for the DAFF) for a given sample.  Due to the
degeneracy of the ground states mentioned above the (staggered)
magnetization of a certain system does not inevitably have a unique
value. Instead, different degenerate ground states of a given system
may have different magnetization values although the energy of the
different ground states is the same, of course. Nevertheless, with our
algorithm we are especially able to find exactly the maximum and the
minimum value of $m$.

\begin{figure}
  \begin{center}
    \epsfxsize=8cm
    \epsffile{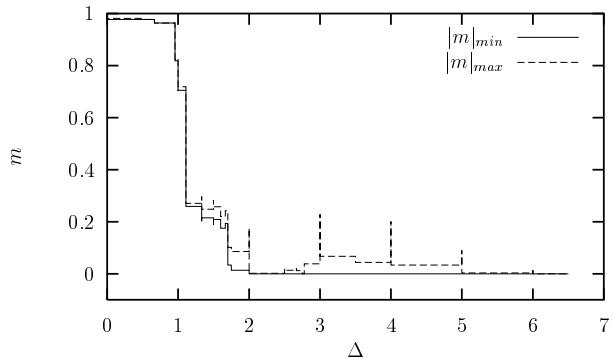}
  \end{center}
  \caption{Maximum and minimum of the absolute value of the staggered
    magnetization of a DAFF ($L=16$, dilution 50\%) versus field.}
  \label{f:example}
\end{figure}

In Figure \ref{f:example} we show the maximum and the minimum absolute
value of the staggered magnetization of one single $L=16$ DAFF sample.
The dilution of the DAFF is 50\%.  As discussed above, $m$ is a
stepwise constant function. It shows strong discontinuities (jumps) at
integer values $n$ of the field $\Delta$.  These jumps are due to the
fact that all the single spins flip at $\Delta = n$ which are
antiparallel to the field and, hence, have a local field of $n$
generated by their $n$ neighbors.  This effect has nothing to do with
the critical behavior but it hinders the scaling analysis. Therefore,
for the scaling analysis of the DAFF we used a higher dilution such
that $\Delta_c$ is well below $1$ and additionally we used larger
lattices sizes such that the critical region is narrow enough so that
all data we need are for values of $\Delta < 1$.

Taking the average over different disorder configurations we can
calculate the order parameter $M = \left[ |m| \right]$ and the
disconnected susceptibility $\chi_{dis} = L^3
\left[m^2\right]$. Here, the square brackets denote an average taken over
up to 180 disorder configuration for the larger system sizes and 6400
disorder configurations for the smaller systems, also depending on how
close the random field strength $\Delta$ is to the critical one.

\begin{figure}
  \begin{center}
    \epsfxsize=8cm
    \epsffile{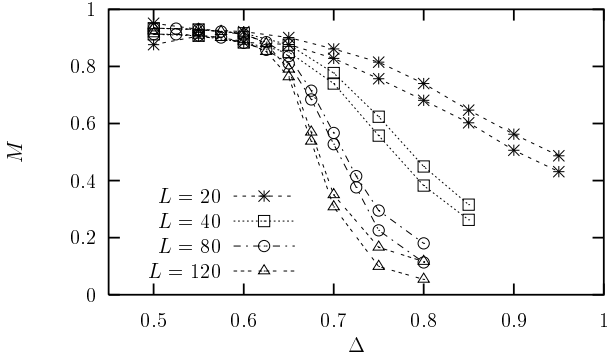}
  \end{center}
  \caption{Averaged maximum and minimum absolute values of the staggered
    magnetization versus field for the DAFF (dilution 55\%, different
    system sizes).}
  \label{f:daff_us}
\end{figure}

For Figure \ref{f:daff_us} we calculated the average of all maximum
values as well the average of all minimum values of $m$.  Although
these values differ significantly we checked that our results do not
depend on whether we take the maximum, the minimum or an average value
of the different occurring (staggered) magnetization values as far as
the scaling behavior of the order parameter is concerned.  Hence, we
decided to neglect the effect of degeneracy and the results presented
here for the order parameter and the disconnected susceptibility are
taken from the maximum values of $m$.  Note, that the considerations
above do not concern the G-RFIM which exhibits a degeneracy of two
only at the jumps, elsewhere it is not degenerate at all.

Apart from that Figure \ref{f:daff_us} demonstrates that for a
dilution of 55\% $\Delta_c$ is well below 1 so that all data we need
for a finite size analysis are smooth functions without steps. 

\begin{figure}
  \begin{center}
    \epsfxsize=8cm
    \epsffile{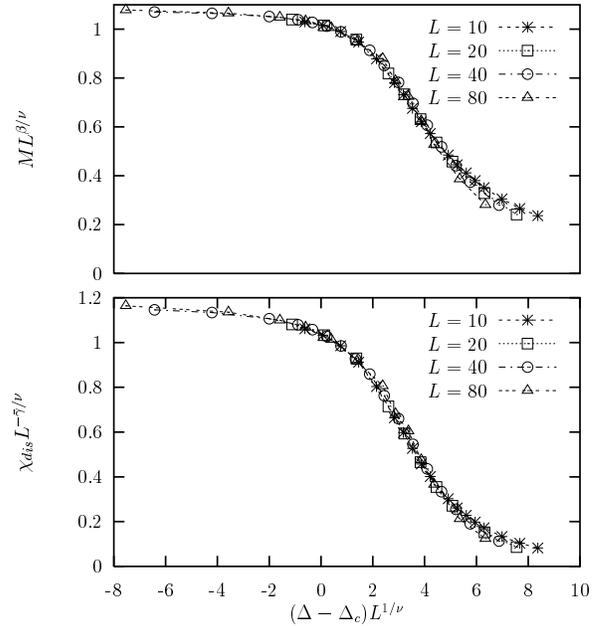}
  \end{center}
  \caption{Scaling plot of the magnetization and the susceptibility
    for the G-RFIM using 
    $\Delta_c = 2.29$, $\nu = 1.19$, $\beta = 0.02$, and $\bar{\gamma}= 3.5$.}
  \label{f:grfim}
\end{figure}

In the following analysis we use the finite size scaling relations
\begin{equation}
  M = L^{-\beta/\nu} \tilde{M}\left((\Delta-\Delta_c)L^{1/\nu}\right)
\end{equation}
for the order parameter and
\begin{equation}
  \chi_{dis} = L^{\bar{\gamma}/\nu}
  \tilde{\chi}\left((\Delta-\Delta_c)L^{1/\nu}\right)
\end{equation}
for the disconnected susceptibility.  Figure \ref{f:grfim} shows the
corresponding scaling plots for the data of the G-RFIM.

Since both quantities, $M$ and $\chi_{dis}$ should have the same critical
field and the same correlation length exponent we adjusted $\Delta_c$
and $\nu$ for both scaling plots at the same time. From this procedure
it follows $\Delta_c = 2.29 \pm 0.04$, $\nu = 1.19 \pm 0.08$, $\beta =
0.02 \pm 0.01$, and $\bar{\gamma} = 3.5 \pm 0.5$.  These are values
which are not surprising and in agreement with most of the previous
work.  The error-bars are estimated from the finite-size scaling.

\begin{figure}
  \begin{center}
    \epsfxsize=8cm
    \epsffile{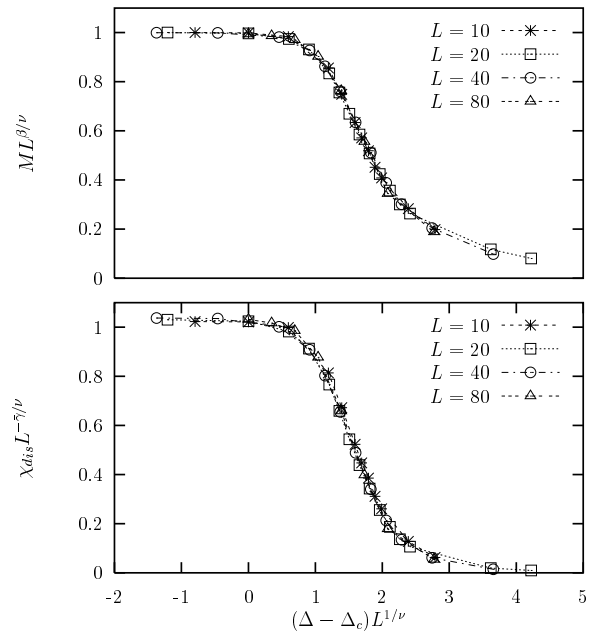}
  \end{center}
  \caption{Scaling plot of the magnetization and the susceptibility
    for the B-RFIM using $\Delta_c = 2.20$, $\nu = 1.67$, $\beta =
    0.0$, and $\bar{\gamma} = 5.0$.}
  \label{f:drfim}
\end{figure}

Figure \ref{f:drfim} shows the same scaling plots for the data of the
B-RFIM.  From the scaling it follows $\Delta_c = 2.20 \pm 0.02$, $\nu
= 1.67 \pm 0.11$, $\beta = 0.0 \pm 0.02$, and $\bar{\gamma} = 5.0 \pm
0.4$.

These values of the critical exponents differ significantly
from those of the G-RFIM suggesting that the two models G-RFIM and
B-RFIM are not in the same universality class. We will discuss this
later in more detail. First we turn to the analysis of the data of the
DAFF.

\begin{figure}
  \begin{center}
    \epsfxsize=8cm
    \epsffile{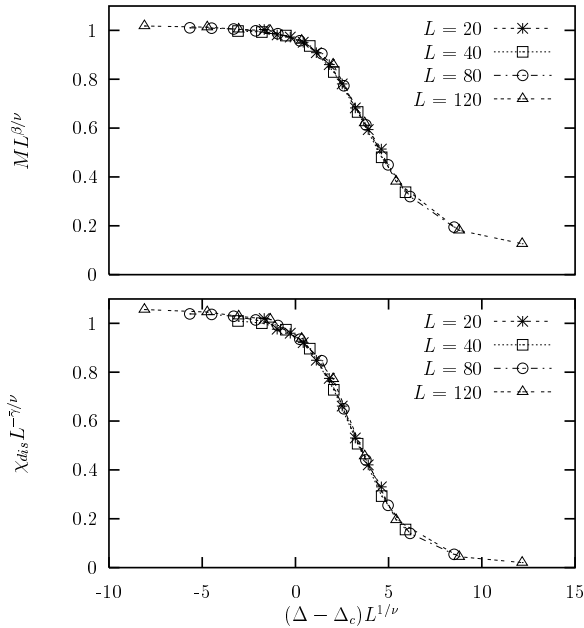}
  \end{center}
  \caption{Scaling plot of the staggered magnetization and the
    corresponding susceptibility for the DAFF using 
    $\Delta_c = 0.62$, $\nu = 1.14$, $\beta = 0.02$, and $\bar{\gamma}
    = 3.4$. The unscaled data are also shown in Figure \ref{f:daff_us}.}
  \label{f:daff}
\end{figure}

Performing the same analysis as before for the data of the DAFF
results in Figure \ref{f:daff}. Here we obtain $\Delta_c = 0.62 \pm
0.03$, $\nu = 1.14 \pm 0.10$, $\beta = 0.02 \pm 0.01$, and
$\bar{\gamma} = 3.4 \pm 0.4$. These values are in good agreement with
the critical exponents we found for the G-RFIM.

\begin{figure}
  \begin{center}
    \epsfxsize=8cm
    \epsffile{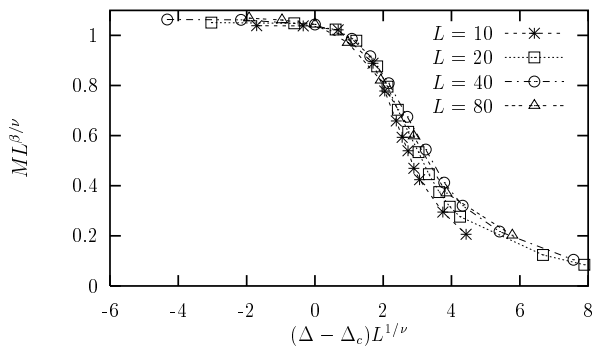}
  \end{center}
  \caption{Scaling plot of the magnetization with the data of the
    B-RFIM but with the exponents of the G-RFIM.
    }
  \label{f:drfim_falsch}
\end{figure}

To prove even more clearly that the data of the B-RFIM cannot be
scaled with the exponents of the DAFF or the G-RFIM we show in Figure
\ref{f:drfim_falsch} data of the B-RFIM which are scaled with the
exponents of the G-RFIM and $\Delta_c$ is best-fitted. Comparing this
Figure with Figure \ref{f:drfim} one can see that the data collapse is
clearly worse.

The following table summarizes all results we extracted from our
finite size scaling analysis.

\begin{center}
  \begin{tabular}{c||c|c|c|c}
     & $\nu$ & $\beta$ &  $\bar{\gamma}$ & $\Delta_c$ \\
    \hline \hline
    G-RFIM    & $1.19 \pm 0.08$  & $0.02 \pm 0.01$ & $3.5 \pm 0.5 $ &
    $2.29 \pm 0.04$ \\ \hline
    B-RFIM & $1.67 \pm 0.11 $ & $0.0 \pm 0.02 $ & $5.0 \pm 0.4 $ &
    $2.20 \pm 0.02$ \\ \hline
    DAFF    & $1.14 \pm 0.10 $ & $0.02 \pm 0.01 $ & $3.4 \pm 0.4 $ &
    $ 0.62 \pm 0.03$\\ 
  \end{tabular}
\end{center}

To summarize, the values we determined for the critical exponents
$\nu$, $\beta $ and $\bar{\gamma}$ of the three-dimensional G-RFIM are
roughly in agreement with the previous numerical works
\cite{ogielski,rieger,newman,esser}. Small deviations -- as far as
they exist -- may be due to the smaller system sizes used in earlier 
numerical investigations or due to the problem of equilibration of
these highly disordered systems in the case of Monte Carlo work.

The values for the critical exponents of the DAFF which we determined
here for the first time within the framework of exact ground-state
calculations agree within the error bars with those of the G-RFIM
confirming that DAFF and G-RFIM belong to the same universality class
\cite{fishman,cardy}. Also, the value of the exponents $\nu$ and
$\bar{\gamma}$ agree reasonably with experimental measurements
\cite{belanger}.

Interestingly the values for the critical exponents of the B-RFIM
deviate from those of G-RFIM and DAFF. This result as well as the
values of $\nu$ are in agreement with previous numerical work
\cite{sourlas,young}.  The fact that $\beta$ is zero may suggest that
the phase transition is of first order as it is the case for the
replica-symmetric mean-field solution \cite{aharony}.  It should be
noted, however, that it was shown by Mezard \cite{mezard} that there
is replica symmetry breaking for the mean field solution of a random-
field model with $m$-component spins in the limit of large $m$.  In
\cite{sourlas} it was also concluded from the exact ground-state
calculations that for the three-dimensional B-RFIM the phase
transition is of first order but on the other hand real space
renormalization yielded deviating results concerning the order of the
phase transition (see e.~g.  \cite{falicov}). Also, the value of $\nu$
is even higher here.  Whether the transition in the D-RFIM is of first
order or not cannot be judged by our simulations.

We should mention that our results are based on a finite-size scaling
analysis and in principle there is the possibility of relevant
logarithmic correction to scaling which possibly could also explain
the deviations of the scaling exponents of the B-RFIM from the
exponents of the G-RFIM and the DAFF. However, such corrections to
scaling are expected for systems at the upper or lower critical
dimension of a system, a case which we do not consider here.

The modified hyperscaling-relation \cite{grinstein} which can be
written in the form $\bar{\gamma} = D \nu -2 \beta$ where D is the
spatial dimension ($D=3$ in our case) is fulfilled by both sets of
exponents.

{\bf Acknowledgments:} The authors acknowledge manifold support by
H.~Horner, G.~Reinelt, and K.~D.~Usadel and fruitful discussions with
N.~Sourlas and J.~-C.~Angl\`{e}s d'Auriac. We are also grateful to the
{\em Paderborn Center for Parallel Computing} for the allocation of
computer time. One of the authors (AH) was supported by the
Graduiertenkolleg ``Modellierung und Wissenschaftliches Rechnen in
Mathematik und Naturwissenschaften'' at the {\em
  In\-ter\-diszi\-pli\-n\"a\-res Zentrum f\"ur Wissenschaftliches
  Rechnen} in Heidelberg and the other (UN) by the Graduiertenkolleg
"Heterogene Systeme" at the {\em Gerhard-Mercator-Universit\"{a}t
  Duisburg}.


\begin{references}
\bibitem{imbrie} J.~Z.~Imbrie, Phys.~Rev.~Lett.~{\bf 53}, 1747 (1984).
\bibitem{bricmont} J.~Bricmont and A.~Kupiainen, Phys.~Rev.~Lett.~{\bf
  59}, 1829 (1987).
\bibitem{imry} Y.~Imry and S.~Ma, Phys.~Rev.~Lett.~{\bf 35}, 1399 (1975).
\bibitem{esser} J.~Esser and U.~Nowak, Phys.~Rev.~B {\bf 55}, 5866 (1997).
\bibitem{bray} A.~J.~Bray and M.~A.~Moore, J.~Phys.~C {\bf 18}, L927 (1985).
\bibitem{nattermann} T.~Nattermann, Phys.~Stat.~Sol. (b) {\bf 131}, 563 (1985).
\bibitem{villain} J.~Villain, J.~Physique {\bf 46}, 1843 (1985).
\bibitem{nowak} U.~Nowak, K.~D.~Usadel and J.~Esser, Physica A {\bf
   250}, 1 (1998).
\bibitem{aharony} A.~Aharony, Phys.~Rev.~B {\bf 18}, 3318 (1978).
\bibitem{swift1} M.~R.~Swift, A.~Martian, M.~Cieplak, and
  J.~R.~Banavar, J.~Phys.~A {\bf 27}, 1525 (1994).
\bibitem{swift2} M.~R.~Swift, A.~J.~Bray, A.~Martian, M.~Cieplak, and
  J.~R.~Banavar, Europhys.~Lett. {\bf 38}, 273 (1997).
\bibitem{sourlas} J.-C.~Angl\`{e}s d'Auriac and N.~Sourlas,
  Europhys.~Lett.~{\bf 39}, 473 (1997).
\bibitem{kleemann} W.~Kleemann, J.~Mod.~Phys.~B {\bf 7}, 2469 (1993).
\bibitem{belanger} D.~P.~Belanger in {\it "Spin Glasses and Random Fields"},
  edited by A. P. Young, World Scientific (1998).
\bibitem{fishman} S.~Fishman and A.~Aharony, J.~Phys.~{\bf C12}, L729 (1979).
\bibitem{cardy} J.~L.~Cardy, Phys.~Rev.~B {\bf 29}, 505 (1984).
\bibitem{nowak2}  U.~Nowak and K.~D.~Usadel, Phys.~Rev.~B {\bf 44},
  7426 (1991).
\bibitem{berker} A.~N.~Berker and S.~R.~McKay, Phys.~Rev.~B {\bf 33},
  4712 (1986).
\bibitem{swamy} M.~N.~S. Swamy and K. Thulasiraman, {\it "Graphs, Networks and
  Algorithms"}, Wiley, New York (1991). 
\bibitem{claibo} J.~D.~Claiborne, {\it "Mathematical Preliminaries for
  Computer Networking"}, Wiley, New York (1990).
\bibitem{knoedel} W.~Kn\"odel, {\it "Graphentheoretische Methoden und ihre
  Anwendung"}, Springer, Berlin (1969).
\bibitem{picard1} J.-C.~Picard and H.~D.~Ratliff, Networks {\bf 5},
  357 (1975).
\bibitem{traeff} J.~L.~Tr\"aff, Eur.~J.~Oper.~Res. 89, 564 (1996).
\bibitem{tarjan} R.~E.~Tarjan, {\it "Data Structures and Network
    Algorithms"}, Society for industrial and applied mathematics,
  Philadelphia (1983).
\bibitem{simone95} C. De Simone, M. Diehl, M. J\"unger, P. Mutzel, 
  G. Reinelt and G. Rinaldi, J. Stat. Phys. {\bf 80}, 487 (1995).
\bibitem{simone96} C. De Simone, M. Diehl, M. J\"unger, P. Mutzel, 
  G. Reinelt and G. Rinaldi, J. Stat. Phys. {\bf 84}, 1363 (1996).
\bibitem{barahona82b} F. Barahona, R. Maynard, R. Rammal and
  J.P. Uhry, J. Phys. A {\bf 15}, 673 (1982).
\bibitem{alex2} A.K. Hartmann, Physica A, {\bf 224}, 480 (1996).
\bibitem{michal92} Z. Michalewicz, {\it "Genetic Algorithms + Data Structures
  = Evolution Programs"}, Springer, Berlin (1992).
\bibitem{pal96} K.F. P\'al, Physica A {\bf 223}, 283 (1996).
\bibitem{alex_stiff} A.K. Hartmann, submitted to Phys.~Rev.~Lett. and
  cond-mat/9806114.
\bibitem{alex_sg2} A.K. Hartmann, Europhys. Lett. {\bf 40}, 429 (1997).
\bibitem{picard2} J.-C.~Picard and M.~Queyranne, Math.~Prog.~Study 13,
  8 (1980).
\bibitem{alex_daff2} A.~K.~Hartmann, Physica A {\bf 248}, 1 (1998).
\bibitem{ogielski} A.~T.~Ogielski, Phys.~Rev.~Lett.~{\bf 57}, 1251 (1986).
\bibitem{rieger} H.~Rieger, Phys.~Rev.~B {\bf 52}, 6659 (1995).
\bibitem{newman} M.~E.~J.~Newman and G.~T.~Barkema, Phys.~Rev.~B {\bf
    53}, 393 (1996).
\bibitem{young} H.~Rieger and A.~P.~Young, J.~Phys. A {\bf 26}, 5279 (1993).
\bibitem{mezard} M.~Mezard and A.~P.~Young, Europhys.~Lett. {\bf 18},
  653 (1992).
\bibitem{falicov} A.~Falicov, A.~N.~Berker, S.~R.~McKay, Phys.~Rev.~B {\bf
  51}, 8266 (1995).
\bibitem{grinstein} G.~Grinstein, Phys.~Rev.~Lett. {\bf 43}, 944 (1976).
\end{references}
\end{document}